\documentclass[aps, prl, reprint, superscriptaddress]{revtex4-1}

\usepackage{amsmath}
\usepackage{amssymb}
\usepackage{wasysym}
\usepackage{graphicx}
\usepackage{hyperref}
\usepackage{color}
\usepackage{physics}
\usepackage{siunitx}
\usepackage{xcolor}
\usepackage{changes}
\usepackage{comment}
\usepackage{siunitx}
\usepackage{bm}

\newcommand*\diff{\mathop{}\!\mathrm{d}}

\providecommand*{\deriv}[3][]{%
\frac{\diff^{#1}#2}{\diff #3^{#1}}}

\allowdisplaybreaks

\begin{document}

\title{Suppression of Unitary Three-body Loss in a Degenerate Bose--Fermi Mixture}

\author{Xing-Yan~Chen}
\thanks{These two authors contributed equally.}
\author{Marcel~Duda}
\thanks{These two authors contributed equally.}
\author{Andreas~Schindewolf}
\author{Roman~Bause}
\affiliation{Max-Planck-Institut f\"{u}r Quantenoptik, 85748 Garching, Germany}
\affiliation{Munich Center for Quantum Science and Technology, 80799 M\"{u}nchen, Germany}
\author{Immanuel~Bloch}
\affiliation{Max-Planck-Institut f\"{u}r Quantenoptik, 85748 Garching, Germany}
\affiliation{Munich Center for Quantum Science and Technology, 80799 M\"{u}nchen, Germany}
\affiliation{Fakult\"{a}t f\"{u}r Physik, Ludwig-Maximilians-Universit\"{a}t, 80799 M\"{u}nchen, Germany}
\author{Xin-Yu~Luo}
\email[]{E-Mail: xinyu.luo@mpq.mpg.de}
\affiliation{Max-Planck-Institut f\"{u}r Quantenoptik, 85748 Garching, Germany}
\affiliation{Munich Center for Quantum Science and Technology, 80799 M\"{u}nchen, Germany}

\date{\today}

\begin{abstract}	
	We study three-body loss in an ultracold mixture of a thermal Bose gas and a degenerate Fermi gas. We find that at unitarity, where the interspecies scattering length diverges, the usual inverse-square temperature scaling of the three-body loss found in non-degenerate systems is strongly modified and reduced with the increasing degeneracy of the Fermi gas. While the reduction of loss is qualitatively explained within the few-body scattering framework, a remaining suppression provides evidence for the long-range RKKY interactions mediated by fermions between bosons. Our model based on RKKY interactions quantitatively reproduces the data without free parameters, and predicts one order of magnitude reduction of the three-body loss coefficient in the deeply Fermi-degenerate regime.
	
\end{abstract}

\maketitle

Ultracold mixtures of bosonic and fermionic atoms provide a powerful platform to explore the physics of Bose--Fermi mixtures. Degenerate mixtures have been produced to investigate phase separation~\cite{Lous2018}, superfluidity~\cite{Ferrier-Barbut2014}, polarons~\cite{Hu2016,Yan2020,Fritsche2021}, and fermion-mediated interactions~\cite{Desalvo2019,Edri2020}. 
Although various phases have been predicted for strongly-interacting mixtures, ranging from supersolid charge density wave states~\cite{Shelykh2010,Cotlet2016} to boson-mediated $s/p$-wave fermion pairing~\cite{Enss2009,Laussy2010,Kinnunen2018}, experimental investigation is hindered by the strong three-body recombination loss between the atoms~\cite{Ospelkaus2006,Gunter2006,Bloom2013,Laurent2017}. Characterizing and understanding the three-body loss is a crucial step towards exploring many-body physics in strongly-interacting Bose--Fermi mixtures.  

Three-body loss describes the process in which two atoms form a dimer while interacting with a third atom. The released binding energy of the dimer leads to the scattering products escaping the trap~\cite{Eismann2016} and to heating~\cite{Weber2003}. A formalism for three-body loss in non-degenerate mixtures has been developed~\cite{Helfrich2010,Petrov2015} and confirmed experimentally~\cite{Bloom2013,Wacker2016,Ulmanis2016}. In the universal regime, where the s-wave scattering length $a$ is much shorter than the de-Broglie wavelength, the three-body loss coefficient $L_3$ is proportional to $a^4$, and can be altered by a series of Efimov resonances arising from couplings to Efimov-trimer bound states. In the unitary regime, where the scattering length is larger than the de-Broglie wavelength, the three-body loss coefficient saturates with $L_3 \propto 1/T^2$~\cite{Petrov2015}, which has been confirmed experimentally in non-degenerate systems~\cite{Ulmanis2016,Eismann2016}. 

In the quantum degenerate regime, the three-body loss rate strongly depends on quantum statistics and other many-body effects such as the fermion-mediated interactions~\cite{De2014,Desalvo2019,Edri2020}. While the three-body recombinations involving identical particles are enhanced (suppressed) by bunching (anti-bunching) due to Bose~\cite{Burt1997,Soding1999,Haller2011} (Fermi~\cite{Ottenstein2008,Huckans2009}) statistics, it remains unexplored how Fermi statistics modifies three-body recombination processes that involve only one fermion. Moreover, the effective boson-boson interaction mediated by the degenerate Fermi gas further complicates the problem. Through the Ruderman--Kittel--Kasuya--Yosida (RKKY) mechanism, two bosons obtain an effective long-range interaction by exchanging one fermion~\cite{Ruderman1954}. The RKKY interaction modifies the scattering potential, and thus, the three-body loss rate. The RKKY interaction is predicted to form the basis for several new quantum phases~\cite{Buechler2003,De2014}, however, so far only mean-field effects of this interaction have been observed~\cite{Desalvo2019,Edri2020}. 

In this Letter, we study three-body loss in a mixture of thermal $^{23}$Na and Fermi-degenerate $^{40}$K, where we explore the effects of both Fermi statistics and the RKKY interaction. We measure the three-body loss coefficient $L_3$ at different interspecies scattering lengths. We find that the loss is described by the zero-range theory in the universal regime, while it is reduced by Fermi degeneracy in the unitary regime. In addition to the $1/T^2$ scaling, the unitary three-body loss decreases with $T/T_F$ of the Fermi gas where $T_F$ is the Fermi temperature. A theoretical model based on few-body scattering theory, including contributions from the Fermi statistics and the RKKY interaction, quantitatively described the data without any free parameters. Based on this model, more than one order of magnitude reduction in $L_3$ can be achieved with $T/T_F < 0.13$.

\begin{figure}
	\centering
	\includegraphics[width=8.5cm]{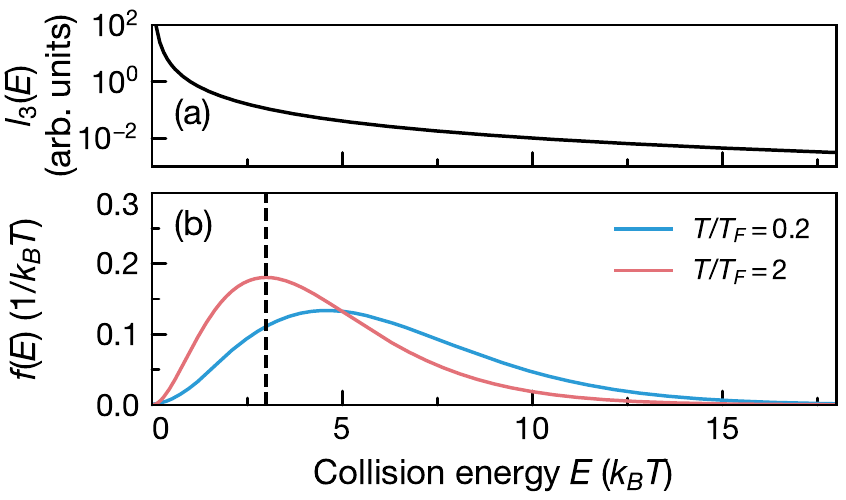}
	\caption{Three-body loss coefficient and collisional energy distribution. (a) Unitary three-body loss coefficient $l_3(E) \propto 1/E^2$, as given by Eq.~\eqref{eq:l3k}. (b) The distribution function $f(E)$ of collisional energy in a Bose--Fermi mixture. We show two scenarios with the same temperature $T$ but different $T_F$. For a thermal mixture (red solid line), the distribution follows the six-dimensional Boltzmann distribution with an average collision energy of $3k_BT$ (black dashed line). For a mixture with a degenerate Fermi gas (blue solid line), the distribution is shifted towards larger collision energies by the Fermi pressure.}
	\label{fig:fig1}
\end{figure}

The reduction of three-body loss is qualitatively explained by the few-body scattering theory. In the unitary regime, instead of the divergent scattering length, the de-Broglie wavelength determines the scattering properties. The unitary three-body loss coefficient for scattering between two identical bosons of mass $m_b$ and one fermion with mass $m_f$ is then given by
\begin{equation}\label{eq:l3k}
	l_3(E) = \frac{8\pi^2\hbar^4\cos^3\phi}{m_r^3 E^2}(1-e^{-4\eta}),
\end{equation}
where $E$ is the kinetic energy in the three-body center-of-mass frame~\cite{Greene2004,Petrov2015}. Here $m_r = m_b m_f/(m_b+m_f)$ is the reduced mass, and $\phi$ is defined by $\sin\phi = m_f/(m_b+m_f)$. The term $1-e^{-4 \eta}$, where $\eta$ is the inelasticity parameter, gives the probability that the incoming wave is not reflected. The average loss coefficient in an atomic mixture is obtained by averaging over the collision energy distributions $f(E)$~\cite{supplement}, 
\begin{equation}\label{eq:fe}
	L_3 = \int  l_3(E) f(E) \diff E.
\end{equation}
For a non-degenerate mixture, the average collision energy is given by $3\,k_BT$, thus $L_3 \propto 1/T^2$ in the unitary regime according to Eq.~\eqref{eq:l3k}. For a mixture where the Fermi gas is degenerate, the average collision energy furthermore depends on the Fermi energy. Due to Fermi statistics, identical fermions distribute over higher momentum states than for the case of a Boltzmann distribution, leading to a larger collision energy as illustrated in Fig.~\ref{fig:fig1}. Accordingly, the average unitary three-body loss decreases as the Fermi energy increases. In other words, the saturation of kinetic energy in the Fermi degenerate regime leads to a reduction of three-body loss compare to thermal gases. To explicitly show the effect of Fermi degeneracy, we separate the $1/T^2$ dependence by defining the temperature-independent loss coefficient $\Lambda \equiv L_3T^2$. In essence, one expects $\Lambda$ to stay constant in the non-degenerate regime and to decrease with $T/T_F$ in the Fermi-degenerate regime.

Our experimental sequence begins with the preparation of a trapped mixture of bosonic $^{23}$Na atoms in $|F,m_F\rangle = |1,1\rangle$ and fermionic $^{40}$K atoms in $|9/2,-9/2\rangle$. Here, $F$ is the total angular momentum and $m_F$ is its $z$-component. The trapping frequencies for Na and K in the $(x,y,z)$-direction are $2 \pi \times (88,141,357)~\mathrm{Hz}$ and $2 \pi \times (97,164,410)~\mathrm{Hz}$, respectively. The interspecies scattering length is varied by tuning the magnetic field around a Feshbach resonance at $\SI{78.30(4)}{G}$~\cite{supplement}. The relation between the scattering length and the magnetic field is obtained by fitting the binding-energy data of the Feshbach molecules, as described in detail in the supplemental materials~\cite{supplement}. To probe the loss on the repulsive (attractive) side of the Feshbach resonance, we prepare the sample at a weakly repulsive interaction below (above) the resonance and ramp the magnetic field in about 100 $\mu$s to the target magnetic field. Before the ramp, a magnetic field gradient is turned on to compensate gravitational sag between the atomic species and ensure good density overlap. After a variable hold time, the magnetic field is ramped back within \SI{100}{\mu s} to a zero-crossing of the interspecies scattering length close the the initial magnetic field. Subsequently the atoms are released from the trap and both species are imaged after some time of flight. We obtain the temperatures and atom numbers from the images and deduce $T_F$ from the atom number and trapping frequencies.

\begin{figure}
	\centering
	\includegraphics[width=8.5cm]{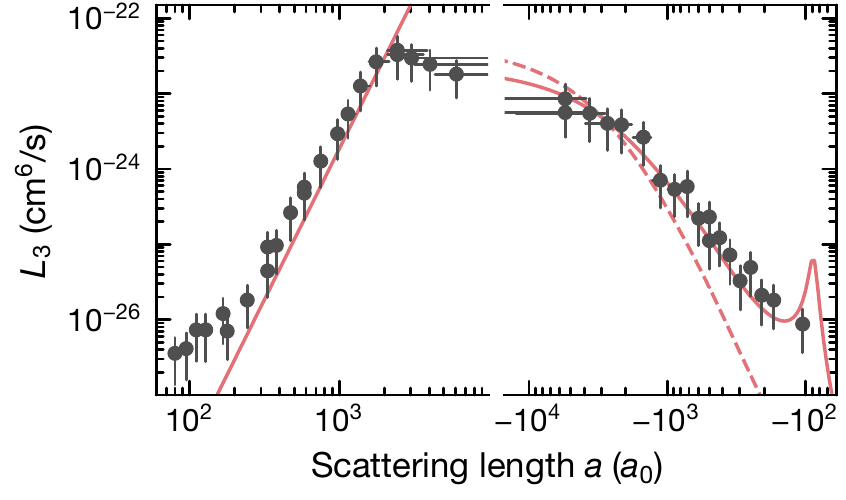}
	\caption{Three-body loss coefficient $L_3$ versus interspecies scattering length $a$ (gray points). The red solid line shows the numerical result of the fitted zero-range theory, which yields an inelasticity parameter $\eta = 0.02$ and a three-body parameter $R_0 = 35\,a_0$ (see the supplemental material~\cite{supplement}). The red dashed line shows the result with $\eta = 0.02$ but without Efimov resonances. The error bars are discussed in the supplement~\cite{supplement}.}
	\label{fig:fig2}
\end{figure}

To characterize the few-body aspect of the three-body loss in our system, we measure $L_3$ at various scattering lengths. We use $ 3 \times 10^5$ Na atoms and $1.5 \times 10^5$ K atoms. The temperature is chosen to be around $0.6\,T_F$ but above the condensation temperature of Na. The measured atom loss ratio between Na and K is close to 2:1, confirming that Na-Na-K is the dominant loss channel, while the K-K-Na three-body loss is suppressed by Pauli blocking between K atoms. We determine $L_3$ by fitting the loss rates of Na and K atoms to the coupled differential equations
\begin{equation}
	\label{eq.dNdt}
	\deriv{N_\mathrm{K}}{t} = \frac{1}{2}\deriv{N_\mathrm{Na}}{t} = -L_3\int n^2_\mathrm{Na}(\bm{x})n_\mathrm{K}(\bm{x}) \diff^3\bm{x}.
\end{equation}
We use a thermal distribution for $n_\mathrm{Na}(\bm{x})$ and a Thomas--Fermi distribution for $n_\mathrm{K}(\bm{x})$. Besides the three-body loss term, we include secondary processes and evaporation in the universal regime~\cite{supplement}. In the unitary regime, these processes become insignificant. 

Fig.~\ref{fig:fig2} summarizes the results of the three-body loss coefficient $L_3$ as a function of the interspecies scattering length $a$. We compare our results to the zero-range theory, which assumes contact interactions~\cite{Petrov2015,Helfrich2010,supplement}. The zero-range theory including finite temperature effects requires averaging over the collision energy distribution, and is not available in an analytic form for $a > 0$. Therefore, we use the zero-temperature formula for $a>0$~\cite{Helfrich2010} and the finite-temperature formula for $a<0$~\cite{Petrov2015}. We find that $\eta \approx 0.02$ and the three-body parameter $R_0 = 35\,a_0$~\cite{supplement} reproduce the loss in the universal regime on both sides of the resonance. In the short range where the scattering length is comparable to the van-der-Waals length $R_\mathrm{vdW} = 53.3\,a_0$, the zero-range approximation breaks down~\cite{Wang2014,Langmack2018,Pricoupenko2019} and the theory fails to describe the data.

With a good understanding of the three-body loss in the universal regime, we move on to probe the unitary three-body loss as a function of temperature and Fermi degeneracy. We use the same experimental sequence as in the previous measurements and fix the probe magnetic field to the pole of the Feshbach resonance $|1/a| \lesssim 10^{-4}{a_0^{-1}}$~\cite{supplement}. We vary $T/T_F$ by changing the initial number of K atoms and the temperature while keeping the mixture in thermal equilibrium. In order to achieve the lowest possible $T/T_F$, we use a high number of K atoms ($\sim 4 \times 10^5$) and a low number of Na atoms ($\sim 3 \times 10^4$). Thus, the loss fraction of K atoms is small compared to the loss fraction of Na atoms and $T/T_F$ is modified by less than 10\% throughout a loss measurement. In the high-$T/T_F$ regime, we reduce the number of K atoms down to $\sim 2 \times 10^4$. Since a dependence of the three-body loss on $T/T_F$ is not expected in this regime, we allow for a relative large increase of $T/T_F$ of about 30\%. The initial temperatures and the trap parameters are chosen such that temperature changes and evaporation are negligible.

\begin{figure}
	\centering
	\includegraphics[width=8.5cm]{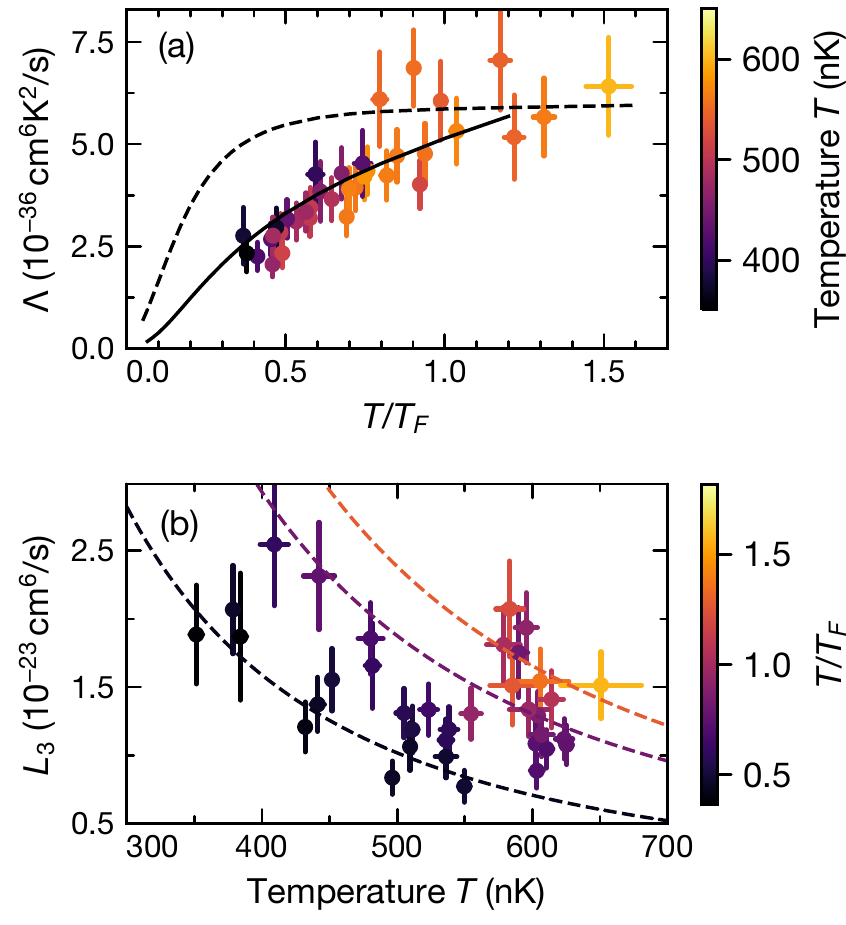}
	\caption{Suppression of unitary three-body loss. (a) Temperature-independent loss coefficient $\Lambda$ as a function of $T/T_F$. The temperature of each loss measurement is indicated by the color of the data points. The dashed line represents the few-body prediction of $\Lambda$ according to Eq.~\eqref{eq.lambda}. The solid line shows the prediction including the RKKY effect. (b) Three-body loss coefficient as a function of temperature. The $T/T_F$ of the Fermi gas is indicated by the color. The dashed lines show the temperature dependence of $L_3 = \Lambda/T^2$ for different $T/T_F$. The $\Lambda$ coefficients are obtained by averaging over the data where $T/T_F$ deviate less than 15$\%$ from $T/T_F = $ 0.4 (black), 0.8 (purple), 1.2 (orange). The error bars are discussed in the supplement~\cite{supplement}.}
	\label{fig:fig3}
\end{figure}

The unitary three-body loss is plotted in Fig.~\ref{fig:fig3}. As shown in Fig.~\ref{fig:fig3}a, the temperature-independent loss coefficient $\Lambda$ is consistent with a saturation for $T/T_F \gtrsim 1$ and decreases with $T/T_F$. A reduction of $T/T_F$ down to 0.4 leads to a reduction of $\Lambda$ by a factor 2.4(4) in comparison to the non-degenerate regime. In order to verify that the reduction does not result from a reduced absolute temperature, the measurements in the same $T/T_F$ regime are taken with different temperatures and atom numbers. As shown in Fig.~\ref{fig:fig3}b, the data for a given $T/T_F$ follow the inverse-square temperature scaling, while for a given temperature $L_3$ decreases with increasing $T_F$ for low $T/T_F$. 

We compare the data with the prediction from the zero-range theory. We use the local density approximation, which treats the mixture at each spatial coordinate in the trap as a homogeneous gas with temperature $T$ and fugacity $z$. The averaged $\Lambda$ over the three-body density overlap is given by
\begin{equation}\label{eq.lambda}
	\Lambda_{f}(T/T_F) = \frac{\int n^2_\mathrm{Na}(\bm{x})  n_\mathrm{K}(\bm{x})  \lambda(z(\bm{x})) \diff^3\bm{x}}{\int n^2_\mathrm{Na}(\bm{x}) n_\mathrm{K}(\bm{x}) \diff^3\bm{x}},
\end{equation}
where the subscript $f$ refers to few-body theory.
Here $z(\bm{x})$ is the fugacity under local density approximation, $\lambda(z) = L_3(z) T^2$ is the local reduced loss coefficient where $L_3(z)$ is given by Eq.~\eqref{eq:fe} in the degenerate regime and is a function of the local fugacity. The increased collision energy by the Fermi pressure leads to a continuous decrease of $\Lambda_f$ in the degenerate regime. While the model shows a similar qualitative dependence as the experiment, the experiment data exhibit a substantially larger reduction: the few-body theory suggests significant reduction only for $T/T_F$ less than $0.1$ while the experimental results already show a reduction for $T/T_F \lesssim 1$. 

\begin{figure}
	\centering
	\includegraphics[width=8.5cm]{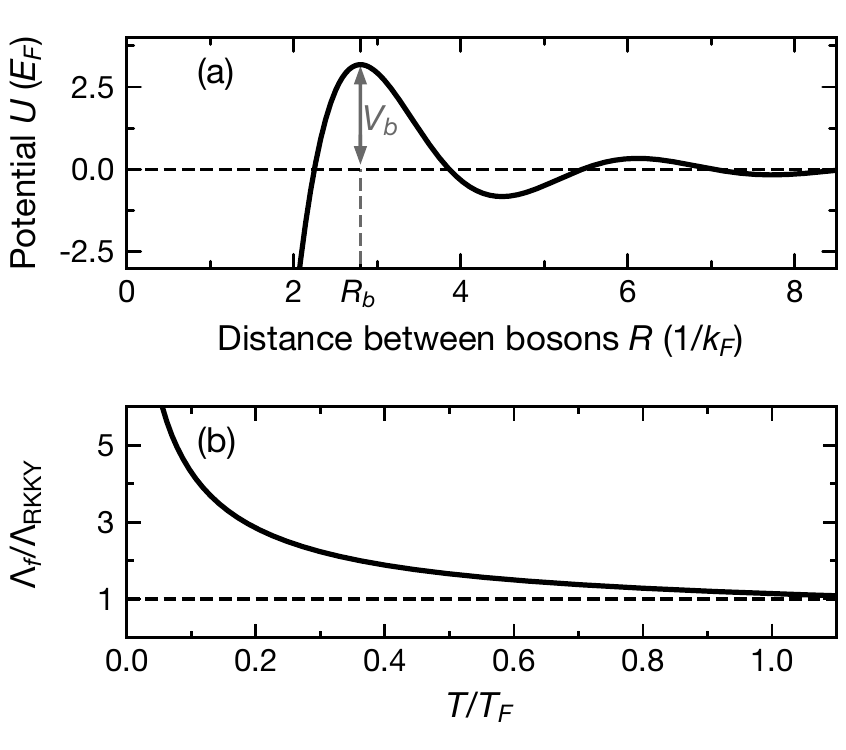}
	\caption{The effect of the RKKY interactions. (a) RKKY potential mediated by fermions between two bosons at unitarity \cite{supplement}. The potential barrier $V_b$ suppresses the tunneling probability into short range. (b) The suppression factor from the RKKY interactions $\Lambda_f/\Lambda_\mathrm{RKKY}$ as a function of $T/T_F$.}
	\label{fig:fig4}
\end{figure}

In the following we show that the fermion-mediated RKKY interactions between bosons are crucial to understand the suppression of the unitary three-body loss. As shown in Fig.~\ref{fig:fig4}a, the RKKY interaction is attractive at short distance and is oscillatory with a length scale $\pi/k_F$ at long range. At a distance $R_b = 2.8/k_F$, the oscillation gives rise to a barrier of the height $V_b \approx 3.2\,E_F \propto T_F$~\cite{supplement}. When the average distance between bosons is much shorter than $R_b$, only the short-range attractive interaction played a role~\cite{Desalvo2019,Edri2020}. In our experiment, the bosons are still thermal with an average distance~$\gtrsim\SI{0.6}{\mu m}$ larger than $R_b \approx \SI{0.3}{\mu m}$. Therefore, the barrier reduces the probability of two bosons approaching each other. In the low-temperature regime, the tunneling probability through the potential barrier $P_T$ is given by the Bethe-Wigner threshold law $\sqrt{E/V_b} \propto \sqrt{T/T_F}$ \cite{quemener2017}, which gives rise to the additional suppression in the degenerate regime. Again, we apply the local density approximation to obtain the local tunneling probability $ P_T(z(\bm{x})) = \sqrt{3k_B T/2V_b(z(\bm{x}))}$ from the potential barrier $V_b(z(\bm{x}))$ and the average kinetic energy of the bosons $3 k_B T/2$. Accordingly, the coefficient $\Lambda$ is given by
\begin{equation}\label{eq.lambda_RKKY}
	\Lambda_\mathrm{RKKY}(T/T_F) = \frac{\int n^2_\mathrm{Na}(\bm{x})  n_\mathrm{K}(\bm{x})  \lambda(z(\bm{x})) P_T(z(\bm{x})) \diff^3\bm{x}}{\int n^2_\mathrm{Na}(\bm{x}) n_\mathrm{K}(\bm{x}) \diff^3\bm{x}}.
\end{equation}
Eq.~\eqref{eq.lambda_RKKY} reproduces the experimental data in the Fermi-degenerate regime without any fitting parameters, as shown in Fig.~\ref{fig:fig3}a. In the deeply degenerate regime, the model predicts more than one order of magnitude reduction, i.e. $\Lambda_\mathrm{RKKY}(T/T_F < 0.13) < 0.1\Lambda_{th}$, where $\Lambda_{th}$ is the reduced loss coefficient in a non-degenerate thermal mixture. The suppression from the RKKY effect can be quantified as $\Lambda_f/\Lambda_\mathrm{RKKY}$, i.e. by comparing to the prediction from the few-body theory. As shown in Fig.~\ref{fig:fig4}b, the suppression factor increase with the Fermi degeneracy. At $T/T_F = 0.13$, where we predict reduction of $\Lambda$ by one order of magnitude compare to $\Lambda_{th}$, the few-body theory predicts a factor of 2.6 reduction, and the RKKY effect suppresses the loss further by a factor of $\Lambda_f/\Lambda_\mathrm{RKKY}\simeq 3.7$. As $T/T_F$ increases, the form of the mediated interaction breaks down due to thermal fluctuations. Therefore we expect a crossover from the prediction with the mediated interactions in the Fermi-degenerate regime to the constant loss in the thermal regime. 

Our model provides a good starting point for further theoretical investigation. Future works could improve the calculation by treating the interaction by non-perturbative methods~\cite{Nishida2009}, and by employing the three-body hyperspherical potential~\cite{DIncao2004,MacNeill2011} to go beyond the Born-Oppenheimer approximation used to derive the mediated interaction.

In conclusion, we have investigated the three-body loss of a Bose--Fermi mixture of $^{23}$Na and $^{40}$K both in the thermal and in the Fermi-degenerate regime. We have confirmed that $L_3$ is proportional to $a^4$ in the universal regime, and is consistent with saturation at unitarity. We have shown that unitary three-body loss is reduced by the Fermi degeneracy, by measuring the temperature-independent loss coefficient $\Lambda$ as a function of $T/T_F$. While the qualitative feature of the reduction is captured by the few-body scattering theory with the degenerate energy distribution, the additional suppression provides strong evidence for the RKKY effect. We have developed a theoretical model based on RKKY interactions, which quantitatively explains the suppression in the degenerate regime without any fitting parameters. Our model predicts a factor of $10$ reduction of $\Lambda$ for $0.13\,T_F$, a very substantial factor that could be reached in a deeply-degenerate mixture. 

The understanding of three-body loss rates in the quantum degenerate regime presented in this work provides a promising outlook to investigate strongly interacting Bose--Fermi mixtures in the deeply degenerate regime. Exciting future works include measuring the unitary collisional loss between a BEC and a degenerate Fermi gas, investigating the universality of unitary Bose--Fermi mixtures~\cite{Ho2004}, probing the Efimov states in the presence of a Fermi sea~\cite{MacNeill2011,Nygaard_2014,Bellotti_2016}, understanding the phase transition from atoms to molecules across the unitary regime~\cite{Bortolotti2008,Bertaina2013,Guidini2015}, and using three-body loss as a tool to probe three-particle correlation functions~\cite{Haller2011}. Our work is also relevant for creating degenerate fermionic molecules from Bose--Fermi mixtures by adiabatically tuning the interaction across the unitary regime~\cite{DeMarco2019,duda2021}, where the suppression of three-body loss could improve the molecule creation efficiency.

\begin{acknowledgments}
	We thank E. Tiemann for providing the coupled-channel calculation of Feshbach molecule binding energy, and D. S. Petrov for providing the theory for atom-dimer loss coefficient and discussion about Efimov resonances. We thank B. Huang, R. Schmidt, and A. Christianen for stimulating discussions. We gratefully acknowledge support from the Max Planck Society, the European Union (PASQuanS Grant No.~817482) and the Deutsche Forschungsgemeinschaft under Germany's Excellence Strategy – EXC-2111 – 390814868 and under Grant No. FOR 2247. A.S. acknowledges funding from the Max Planck Harvard Research Center for Quantum Optics.
\end{acknowledgments}

\bibliography{bibliography2}

\clearpage

\section*{Supplemental Material for "Suppression of Unitary Three-body Loss in a Degenerate Bose--Fermi Mixture"}

\subsection{Radio-frequency spectroscopy of weakly bound Feshbach molecules}\label{rf-spectroscopy}
We determine the magnetic-field dependence of the interspecies scattering length $a$ around the Feshbach resonance at \SI{78.3}{G} by characterizing the binding energy $E_b$ of the Feshbach molecules via radio-frequency spectroscopy on the repulsive side of the resonance. We begin the measurement in a crossed optical dipole trap with an ultracold mixture of about $3\times 10^5$ Na atoms in $|1,1\rangle$ and about $2\times 10^5$ K atoms in $|9/2,-7/2\rangle$. The temperature $T$ of the sample is about \SI{400}{nK}. The trapping frequencies of the Na and K atoms in $(x,y,z)$-direction are $2 \pi\times (74,110,292)~\mathrm{Hz}$ and $2 \pi \times (79,133,333)~\mathrm{Hz}$, respectively.

We apply a radio-frequency pulse at a given frequency~$\nu$. Afterwards, we turn off the trapping light and separate the K atoms in state $|9/2,-9/2\rangle$ from the K atoms in the state $|9/2,-7/2\rangle$ with a magnetic field gradient and detect the number of atoms in both states with absorption imaging. We perform this measurement for various frequencies and various magnetic fields $B$ between \SI{73.5}{G} and \SI{78.0}{G} and obtain the calibrated magnetic field with the Breit--Rabi formula \cite{Tiecke2011}.

The radio-frequency spectroscopy of the weakly bound Feshbach-molecule state $|FB\rangle$ relies on the atom-dimer loss of Feshbach molecules and unbound atoms. The atom-dimer loss depletes the detected number of atoms and allows us to accumulate signal over time when we drive the $|9/2,-7/2\rangle\rightarrow|FB\rangle$ transition, even though the molecule association efficiency is quite low. By increasing the pulse length with increasing binding energy, we ensure that the lost fraction of the Na and K atoms stays larger than $10\%$. After that, we release the atoms from the trap and detect the remaining number of unbound Na and K atoms. 

\begin{figure}
	\centering
	\includegraphics{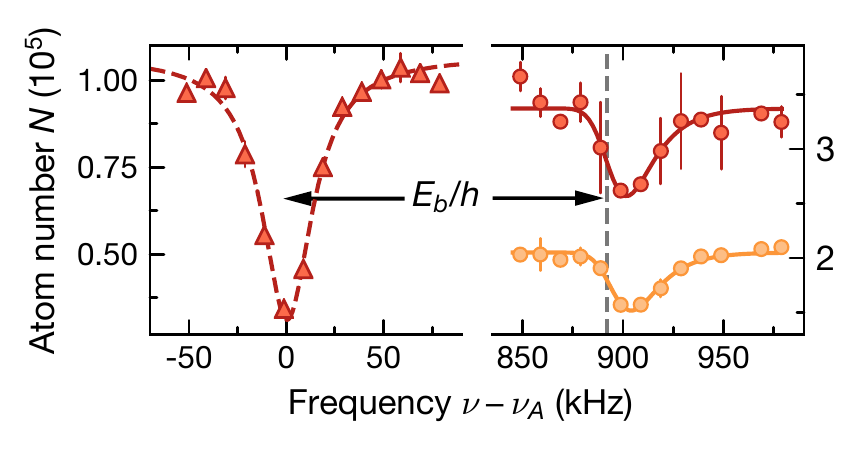}
	\caption{Radio-frequency spectrum for the determination of the binding energy at \SI{76.5}{G}. On the left, the remaining number of K atoms in the $|9/2,-7/2\rangle$ state from radio-frequency transfer (red triangles) to the $|9/2,-9/2\rangle$ state at $\nu_A$ and its fit (red dashed line) are shown. On the right, the remaining numbers of K atoms (red circles) and Na atoms (orange circles) from radio-frequency loss spectroscopy of the weakly bound Feshbach state $|FB\rangle$ and their fits (red and orange solid lines) are shown. The vertical gray dashed line indicates the transition frequency to $|FB\rangle$ extracted from Eq.~\eqref{eq.rf} for atoms at rest. From there the binding energy is given relative to $\nu_A$.}
	\label{fig:rf}
\end{figure}

\begin{figure}
	\centering
	\includegraphics{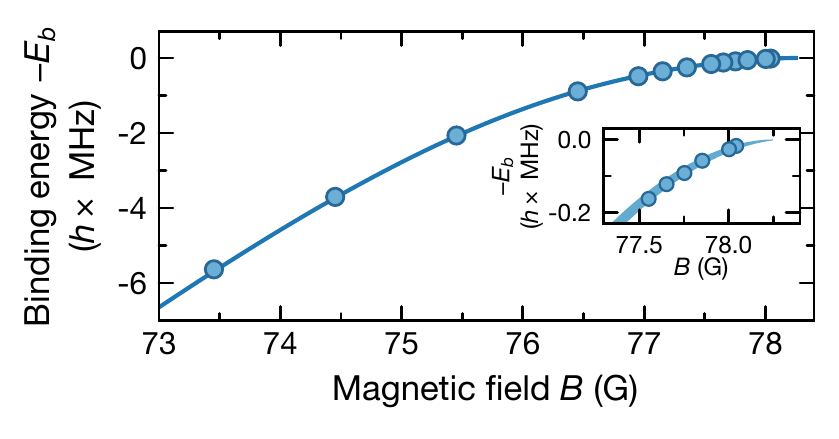}
	\caption{
		Binding energy $E_b$ of the Feshbach-molecule state $|FB\rangle$ as a function of the magnetic field. The circles mark the fit results from the radio-frequency spectra. The solid line shows the function of Eq.~\eqref{eq.Eb} fitted to these data. Inset: Close-up for binding energies below 200 kHz. The uncertainty of the fit function is smaller than the thickness of the solid line in the main figure, however, it is resolved in the close-up. This uncertainty includes statistical errors of the fit and an uncertainty of 5\% of the background scattering length $a_{bg} = -619(31)\,a_0$ \cite{Viel2016}.
	}
	\label{fig:figs2}
\end{figure}

From the radio-frequency spectrum we determine the binding energy of the Feshbach molecules, see Fig.~\ref{fig:rf} as an example. We assume that the number of atoms that get lost in atom-dimer collisions scales proportionally with the number of formed molecules $N_\mathrm{mol}$. In this case, the line shape of the radio-frequency spectrum of the bound state can be modeled via Fermi's golden rule as \cite{Klempt2008}
\begin{equation}\label{eq.rf}
	N_\mathrm{mol}(\nu) \propto \int_{0}^{\infty} \diff \epsilon_r  F(\epsilon_r) h(\epsilon_r) e^{-(h\nu-E_b-h\nu_A-\epsilon_r)^2/\sigma^2}.
\end{equation}
Here, $\nu$ is the radio-frequency, and $h \nu_A$ is the atomic transition energy. The molecule number $N_\mathrm{mol}$ is proportional to the product of $h(\epsilon_r)$, which is the number of colliding pairs per relative kinetic energy interval $\epsilon_r$, and the Franck--Condon factor $F(\epsilon_r)$ between the unbound atom pair and the bound molecular state. The product is convoluted with a Gaussian distribution with the width $\sigma$ to account for the finite energy resolution. We adopt the following simplified Franck--Condon factor $F(\epsilon_r) \propto \sqrt{\epsilon_r}(1+\epsilon_r/E_b)^{-2}$ from Ref.~\cite{Chin2005}. The function $h(\epsilon_r)$ is proportional to the Boltzmann factor $e^{-\epsilon_r/k_B T}$, where the temperature $T$ of the atomic cloud is obtained from time-of-flight images. 
Following Eq.~\eqref{eq.rf}, we fit functions to the lost number of atoms in the radio-frequency spectra of the bound state, as demonstrated in Fig.~\ref{fig:rf}. Note that the relative kinetic energy of the associated atoms has to be transferred into the microwave field and therefore increases the transition frequency. The binding energies $E_b$ that we extract from these fits are presented in Fig.~\ref{fig:figs2} as a function of the magnetic field.

\subsection{Determination of the magnetic-field dependent scattering length}\label{scattering-length}
To determine the scattering length from the binding energy near the interspecies Feshbach resonance, we adopt the model for overlapping Feshbach resonances from Ref.~\cite{Lange2009} to include a resonance at \SI{89.7}{G}. The binding energy is given by solving
\begin{equation}\label{eq.Eb}
	\frac{\sqrt{2m_r E_b}}{\hbar} = \frac{1}{a_\mathrm{bg} - \bar{a}} + \frac{1}{\bar{a}}\sum_{i = 1,2}\frac{\Gamma_i}{E_b+E_i},
\end{equation}
where $E_1$ and $E_2$ are the energies of the bare molecular state for the resonance at \SI{78.3}{G} and \SI{89.7}{G}, respectively, $a_\mathrm{bg}$ is the background scattering length, $m_r$ is the reduced mass, $\bar{a} = 4\pi\Gamma(1/4)^{-2}R_\mathrm{vdW} \approx 51\,a_0$ is the mean scattering length, where $\Gamma(x)$ is the gamma function and $R_\mathrm{vdW}$ is the van-der-Waals length. The energy of the bare molecular states can be tuned magnetically with $E_i = \delta\mu_i(B-B_{c,i})$, where $\delta\mu_i$ is the relative magnetic moment with respect to the dissociation threshold and $B_{c,i}$ is the magnetic field at which the bare molecular state crosses the dissociation threshold. The Feshbach coupling strengths between the open and closed channels are denoted by $\Gamma_i$. For an isolated resonance, the Feshbach coupling strength $\Gamma$ is related to the commonly used Feshbach resonance width $\Delta$ as $\Gamma = 2\Delta\delta\mu\alpha^{-1}$, where $\alpha = (\bar{a}-a_\mathrm{bg})^2/a_\mathrm{bg}\bar{a}$. For overlapping resonances, $\Gamma_i$ cannot be independently extracted without considering nearby resonances. We fit Eq.~\eqref{eq.Eb} to the binding-energy data to extract the parameters for the two resonances. To avoid overfitting, we take $\Gamma_1$ as the only free fit parameter while all other parameters are derived from a coupled-channel calculation \cite{TiemannCommu}. The fitted binding energy is presented in Fig.~\ref{fig:figs2} as function of the magnetic field. The parameters are tabulated in Tab.~\ref{tab.fit}. 

The scattering length $a$ is given by solving \cite{Lange2009}
\begin{equation}\label{eq.aB}
	\frac{1}{a - \bar{a}}= \frac{1}{a_\mathrm{bg} - \bar{a}} + \frac{1}{\bar{a}}\sum_{i = 1,2}\frac{\Gamma_i}{E_i}.
\end{equation}
With the parameters extracted from Eq.~\eqref{eq.Eb}, we can compute the scattering lengths at different magnetic fields using Eq.~\eqref{eq.aB}. We further rewrite the expression of the scattering length as
\begin{equation}
	\frac{a}{a_\mathrm{bg}} = \prod_{i = 1,2}\frac{B-B_i^*}{B-B_{0,i}},
\end{equation}
where we define $B_i^*$ as the $i$-th zero crossing of $a(B)$ and $B_{0,i}$ the $i$-th pole of $a(B)$. We obtain the resonance position $B_{0,1} = \SI{78.30(4)}{G}$ with a width of $\Delta = B_{0,1} - B_1^* = \SI{5.27(4)}{G}$.

\begin{table}
	\centering
	\begin{tabular}{ccccccc}
		\hline\hline
		$i$ &$\Gamma_i/h$ (MHz) & $\delta\mu_i$ $(\mu_B$) & $B_{c,i}$ (G) & $B_i^*$ (G) & $B_{0,i}$  (G)  \\
		\hline
		1 & 4.180(9) & 1.894  & 73.92   & 73.034(4)  & 78.30(4)       \\
		2 & 1.385  & 2.085    & 80.58    & 80.358(1)   & 89.7(6)      \\
		\hline\hline
	\end{tabular}
	\caption{
		Parameters of the Feshbach coupling strengths $\Gamma_i$, the differential magnetic moments $\delta\mu_i$, and the crossing of the bare molecular states with the dissociation threshold $B_{c,i}$ which are used to fit the model for overlapping Feshbach resonances described by Eq.~\eqref{eq.Eb}. $\Gamma_1$ is the only free fit parameter while the others are extracted from a coupled-channel calculation \cite{TiemannCommu}. 
	}\label{tab.fit}
\end{table}

\subsection{Secondary loss and heating}
In the universal regime, we consider in addition to Eq.~(3) the anti-evaporative heating, evaporation, and secondary processes. The extended coupled-differential equations are given by
\begin{align}
	\label{eq.dNadt}
	\deriv{N_\mathrm{Na}}{t} &= -(2+\delta)L_3\int n^2_\mathrm{Na}n_\mathrm{K} \diff^3\bm{x} + \left(\deriv{N_\mathrm{Na}}{t}\right)_\text{ev}, \\
	\label{eq.dKdt}
	\deriv{N_\mathrm{K}}{t} &= -L_3\int n^2_\mathrm{Na}n_\mathrm{K} \diff^3\bm{x}, \\
	\begin{split}
		\label{eq.dTdt}
		\deriv{T}{t} &= \frac{(\frac{3}{2} - \beta) T+\frac{1}{3}T_h}{N_\mathrm{Na}+N_\mathrm{K}}L_3\int n^2_\mathrm{Na}n_\mathrm{K} \diff^3\bm{x} + \left(\deriv{T}{t}\right)_\text{ev}.
	\end{split}
\end{align}
Here, $\beta$ describes the anti-evaporative heating due to the three-body loss, the parameter $\delta$ describes the secondary loss, $T_h$ describes the secondary heating, and $\left(\deriv{N_\mathrm{Na}}{t}\right)_\text{ev}$ and $\left(\deriv{T}{t}\right)_\text{ev}$ give the contributions from evaporative cooling.

Anti-evaporation is caused by the fact that the three-body loss predominantly takes place at the bottom of the trap. Consequently, the averaged potential energy lost per particle is given by
\begin{equation}
	\beta k_B T = \frac{\int n^2_\mathrm{Na}n_\mathrm{K} (2U_\mathrm{Na} + U_\mathrm{K}) \diff^3\bm{x}}{3\int n^2_\mathrm{Na}n_\mathrm{K} \diff^3\bm{x}},
\end{equation}
where $U_\mathrm{Na}$ and $U_\mathrm{K}$ are the trap-depths for the respective species, is smaller than the averaged thermal energy per particle $(3/2)k_B T$. 

\begin{figure}
	\centering
	\includegraphics{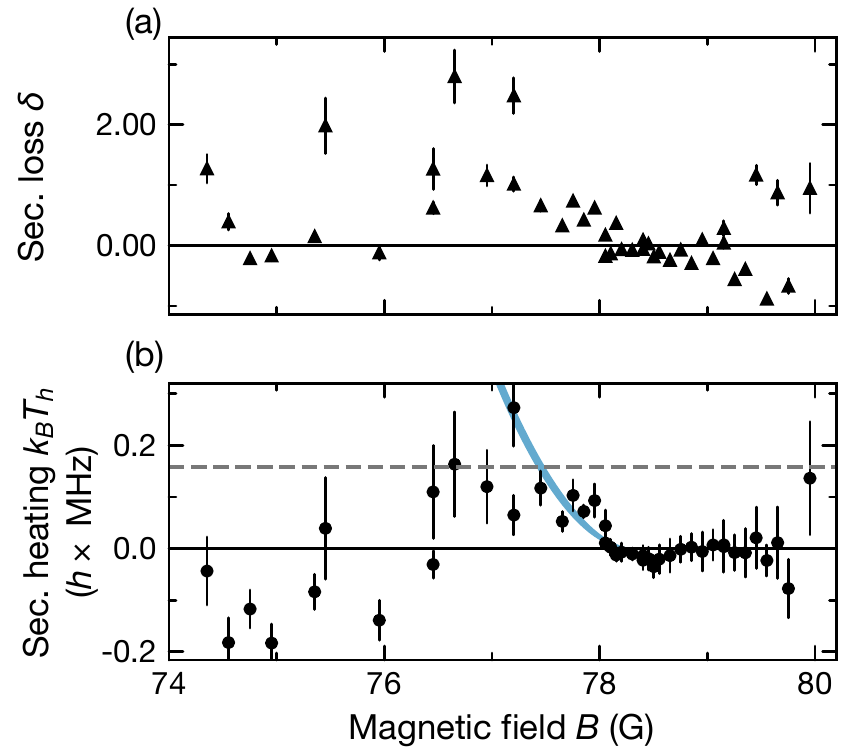}
	\caption{Secondary loss and secondary heating as a function of the magnetic field. (a) The data (black triangles) show the additional number of Na atoms that are in average lost per recombination event due to secondary collisions. (b) The data (black points) show the excess energy deposited in average per recombination event due to secondary collisions. The blue solid line represents the kinetic energy the Na atoms carry away from a three-body collision as given in Eq.~\eqref{eq.secondary_heating} and the gray dashed line is the trap depth of the Na atoms.}
	\label{fig:secondary}
\end{figure}

Secondary loss occurs on the repulsive side of the Feshbach resonance when the released binding energy of the formed dimer is not sufficient to expel the dimer from the trap. The trapped dimer collides with another Na atom in a secondary collision, which causes the dimer to relax into deeply bound states. This releases sufficient energy to expel the products from the trap, leading to additional loss described by $\delta$. Secondary loss predominantly affects Na, as the collisions between the trapped dimer and K atoms are suppressed by Pauli blocking between the K atom and the weakly bound K atom inside the dimer~\cite{Bloom2013}. Secondary heating occurs when the Na atom released from the three-body recombination remains trapped. The trapped Na can thermalize with the cloud and release an average energy of $k_B T_h$. The extracted parameters for the secondary processes are shown in Fig.~\ref{fig:secondary}. 

Our typical trap depth for Feshbach molecules is $h\times\SI{0.5}{MHz}$. For magnetic fields above \SI{75.5}{G}, the binding energy is comparable to the trap depth. A weakly bound molecule formed by three-body recombination cannot escape the trap and can collide with another Na atom, leading to atom-dimer loss. As shown in Fig.~\ref{fig:secondary}(a), the average loss of Na atoms per recombination $\delta$ vanishes above the Feshbach resonance while it increase below the resonance until \SI{75.5}{G}.
In Fig.~\ref{fig:secondary}(b), the secondary heating $T_h$ is compared to the kinetic energy of the Na atom after the recombination, which is given by 
\begin{equation}\label{eq.secondary_heating}
	\frac{m_\mathrm{Na}+m_\mathrm{K}}{2m_\mathrm{Na}+m_\mathrm{K}}E_b.
\end{equation}
We find that $T_h$ qualitatively follows Eq.~\eqref{eq.secondary_heating} until it reaches the trap depth of the Na atoms. For lower magnetic fields $T_h$ decreases again since the Na atom can escape the trap after the recombination process.

\subsection{Evaporation}

The evaporative cooling of Na atoms mitigates the anti-evaporative heating from three-body recombination. Since the Na atoms experience a significantly shallower trap than the K atoms, they sympathetically cool the mixture via evaporation. We consider the evaporation of Na atoms initiated by both Na-Na and Na-K collisions. The evaporation of the Na atoms is then given by
\begin{align}
	\label{eq.na_evap}
	\left(\deriv{N_\mathrm{Na}}{t}\right)_\text{ev} &= - \gamma_1 \frac{N^2_\mathrm{Na}}{T} - \gamma_2 \frac{N_\mathrm{Na}N_\mathrm{K}}{T},\\
	\label{eq.k_evap}
	\left(\deriv{T}{t}\right)_\text{ev} &= -\frac{1}{3}(\xi+\kappa-3)\frac{\gamma_1 N_\mathrm{Na} + \gamma_2 N_\mathrm{K}}{N_\mathrm{Na} + N_\mathrm{K}}, \\
	\gamma_1  &= \frac{8}{\pi}\frac{m_\mathrm{Na}\bar{\omega}_\mathrm{Na}^3}{k_B}a_{bb}^2e^{-\xi}V_r,\\
	\gamma_2  &= \frac{4}{\pi}\frac{m_\mathrm{K}\bar{\omega}_\mathrm{K}^3}{k_B(a^{-2}+ 2m_r U_\mathrm{Na} \hbar^{-2})}e^{-\xi}V_r,\\
	\kappa &= \left( 1-\frac{P(5,\xi)}{P(3,\xi)}\right) V_r, \\
	V_r &= \xi - 4\frac{P(4,\xi)}{P(3,\xi)},
\end{align}
where the first (second) term in Eq.~\eqref{eq.na_evap} accounts for evaporation induced by intraspecies (interspecies) collisions. Here $\xi = U_\mathrm{Na}/k_B T$ is the truncation parameter for Na, $P(a,\xi)$ is the regularized incomplete Gamma function
\begin{equation}
	P(a,\xi)=\frac{\int_0^\xi u^{a-1}e^{-u}{\rm d}u}{\int_0^{\infty}u^{a-1}e^{-u}{\rm d}u},
\end{equation}
and $\bar{\omega}_\mathrm{Na}$ $(\bar{\omega}_\mathrm{K})$ is the geometric mean of the trapping frequencies for Na (K) atoms.

\begin{figure}
	\centering
	\includegraphics{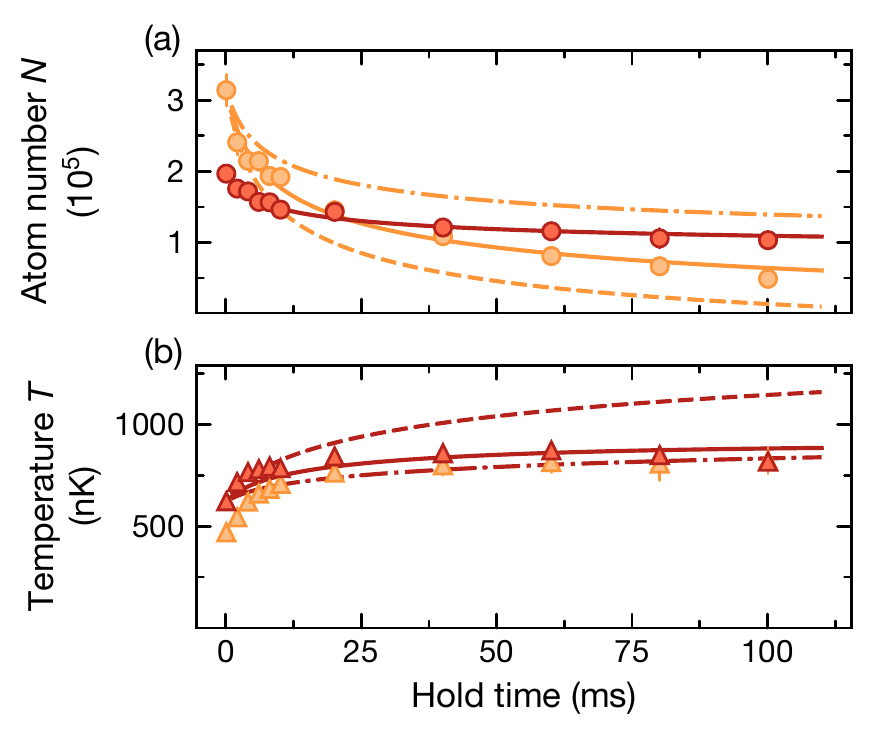}
	\caption{Atom loss and heating at 77.8 G. (a) Atom number of Na (orange circles) and K (red circles) as a function of the hold time. (b) Temperature of Na (orange triangles) and K (red triangles) as a function of the hold time.
		The solid lines show a fit of the coupled differential equations described in Eqs.~\eqref{eq.dNadt}-\eqref{eq.dTdt} using the temperature of the K atoms as temperature of the mixture.
		The dashed (dash-dotted) lines represent the same solution omitting the evaporation (secondary loss and secondary heating processes).
	}
	\label{fig:loss}
\end{figure}
The intraspecies evaporation rate follows Refs.~\cite{Luiten1996,Eismann2016}, by replacing the collision cross section with $\sigma_{bb} = 8\pi a_{bb}^2$, where $a_{bb}$ is the intraspecies scattering length of Na. The interspecies evaporation (i.e., the sympathetic cooling) \cite{Mosk_2001} is obtained by using the Na-K collision cross section $\sigma_{bf} = 4\pi/(a^{-2}+ k^2)$ and replacing the density of Na atoms with the density of the K atoms. Since only collisions with relative kinetic energy larger than the trap depth lead to evaporation \cite{Eismann2016}, we substitute $k \simeq \sqrt{2m_r U_\mathrm{Na}}/\hbar$. to obtain the energy-independent cross section $\sigma_{bf} = 4\pi/(a^{-2}+ 2m_r U_\mathrm{Na}\hbar^{-2})$. 

The contribution of the secondary processes and of evaporation to the decay model are exemplarily illustrated in Fig.~\ref{fig:loss}. Secondary loss and the secondary heating contribute in particular at short hold times where the three-body recombination rate is large. On longer hold times the evaporation kicks in and compensates heating. We separately fit $L_3$ using the temperature of the Na and K atoms in order to estimate the systematic error due to temperature discrepancies between the atomic species.

\subsection{Theory prediction for the three-body loss coefficient}\label{theory1}
We compare the measured $L_3$ to the zero-range theory for heteronuclear mixtures \cite{Helfrich2010,Petrov2015}. For negative scattering lengths, the recombination coefficient at finite temperature in a non-degenerate sample is given by
\begin{multline}
	L_3(a<0) = 4 \, \pi^2 \cos^3 \! \phi \,\, \frac{\hbar^7}{m_r^4 (k_B T)^3}
	\left( 1 - e^{-4 \eta} \right)
	\\
	\times
	\int_0^\infty
	\frac{1 - |s_{11}|^2}{|1 + (k R_0)^{-2 i s_0} e^{-2\eta} \, s_{11} |^2}
	\,e^{-\hbar^2k^2/2m_r k_BT}
	\,
	k \, \diff k,
	\label{eq:l3_T}
\end{multline}
where the angle $\phi$ is defined by $\sin\phi = m_\mathrm{K}/(m_\mathrm{Na}+m_\mathrm{K})$, $R_0$ is the three-body parameter, $\eta$ is the inelastic parameter, $s_0 = 0.285$ for NaK, and $s_{11}$ is a universal function which depends on $ka$ and the mass ratio between the species. Here, $k$ is the three-body collision wave vector which relates to the collision energy as $E_k = \hbar^2k^2/2m_r$.

For positive scattering lengths where temperature averaging is more evolved, we compare our measurement results to the zero-temperature formula \cite{Helfrich2010}
\begin{align}
	L_3(a&>0)=C_\alpha \bigg( \frac{\text{sin}^2[s_0~\text{ln}(a/a_+)]+\text{sinh}^2\eta}{\text{sinh}^2(\pi s_0 + \eta)+\text{cos}^2[s_0~\text{ln}(a/a_+)]}\nonumber\\
	&+\frac{\text{coth}(\pi s_0)\text{cosh}(\eta)\text{sinh}(\eta)}{\text{sinh}^2(\pi s_0 + \eta)+\text{cos}^2[s_0~\text{ln}(a/a_+)]} \bigg) \frac{\hbar a^4}{m_\mathrm{K}},
	\label{eq:l3apos}
\end{align}
where $a_+$ marks the minima in the $L_3$. The regime we measured is away from any Efimov resonances so the term $\cos[s_0~\text{ln}(a/a_+)]$ is set to zero and the term $\sin[s_0~\text{ln}(a/a_+)]$ is set to one.

\subsection{Distribution of the three-body collision energy}
The distribution of the three-body collision energy is given by
\begin{align}
	f(E) &= \int \delta(\epsilon_1 + \epsilon_2 + \epsilon_3 - E_\mathrm{cm} - E) \nonumber \\ 
	& \times f_b(\bm{k_1})f_b(\bm{k_2})f_f(\bm{k_3})\diff^3\bm{k_1}\diff^3\bm{k_2}\diff^3\bm{k_3},
\end{align}
where $\epsilon_1 = \frac{\hbar^2\bm{k_1}^2}{2m_b} $, $\epsilon_2 = \frac{\hbar^2\bm{k_2}^2}{2m_b} $, $\epsilon_3 = \frac{\hbar^2\bm{k_3}^2}{2m_f} $, and $E_\mathrm{cm}= \frac{\hbar^2 (\bm{k_1}+\bm{k_2}+\bm{k_3})^2}{2(2m_b + m_f)}$. We use the Boltzmann distribution $f_b(k) = e^{-\hbar^2 k^2/(2m_b k_B T)}$ for the Bose gas, and the Fermi-Dirac distribution  $f_f(k) = (z^{-1}e^{\hbar^2 k^2/(2m_f k_B T)} + 1)^{-1}$ for the Fermi gas. 

\subsection{Atom--dimer loss}
The two-body loss coefficient between Na atoms and NaK Feshbach molecules provides an alternative way to determine the inelasticity and three-body parameters \cite{Helfrich2010}. We start with an atom mixture and ramp the magnetic field across the Feshbach resonance at a rate of 3.5~G/ms to create Feshbach molecules. To prepare a sample of molecules and Na atom mixture, we use a resonant laser pulse to remove the unpaired K atoms. The clear out is done at 50~G to detune the molecular transition from the laser. We then ramp to different magnetic fields near the Feshbach resonance and measure the atom and molecule loss over the holding time. The remaining atoms and molecules are detected after time-of-fight expansion in the presence a magnetic field gradient. The molecules are dissociated before imaging by ramping the magnetic field back across the Feshbach resonance. The Na + NaK loss coefficient $\beta_\mathrm{Na+NaK}$ is extracted by the differential equation
\begin{equation}
	\deriv{N_\mathrm{NaK}}{t} = \deriv{N_\mathrm{Na}}{t} = -\beta_\mathrm{Na+NaK}\int n_\mathrm{Na}n_\mathrm{NaK} \diff^3\bm{x}.
\end{equation}
The measurement is typically performed at 250~nK without significant heating throughout the hold time.

\begin{figure}
	\centering
	\includegraphics{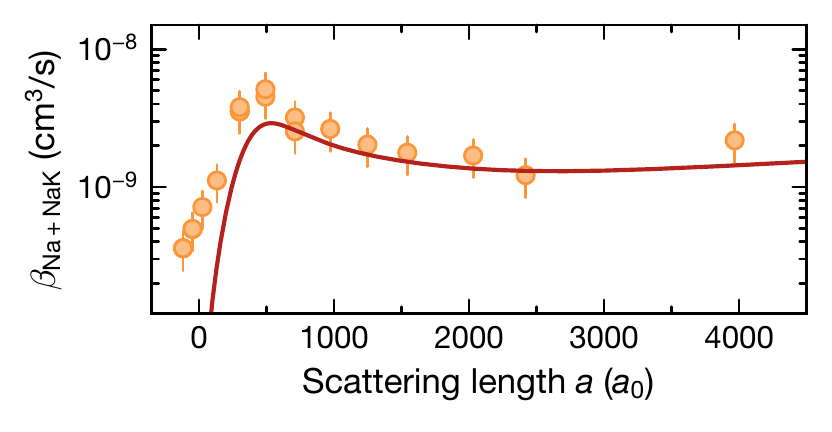}
	\caption{Na + NaK loss coefficient vs. scattering length (orange circles). The red solid line shows the zero-range theory with $\eta = 0.1$ and $R_0 = 200a_0$ \cite{PetrovCommu}.}
	\label{fig:figS4}
\end{figure}

The extracted loss coefficient and the theory fit are shown in Fig.~\ref{fig:figS4}. The resonance at $a \approx 500\,a_0$ in $\beta_\mathrm{Na+NaK}$ is a clear signature of an Efimov state, from which we determine $\eta = 0.1$ and $R_0 = 200\,a_0$ using the zero-range theory with temperature averaging \cite{Helfrich2010}. Note that the Efimov resonance here belongs to a different Efimov state than the resonance determined from the three-body loss data in Fig.~{2} in the main text. The difference between the $R_0$ extracted from three-body and atom-dimer loss is within a factor of 6, which is negligible compared to the ratio between the scattering lengths of neighboring Efimov resonances, which is given by the Efimov scaling factor ($e^{\pi/s_0}\simeq6\times10^4$). The difference between $R_0$ on different side of the Feshbach resonance could potentially be explained by the formalism in \cite{Huang2014}.

\subsection{Interspecies interaction strength in the fermion-mediated interactions}

The mediated interaction is given by \cite{Ruderman1954,De2014}
\begin{equation}
	U(R) = -\frac{2 m_f g^2 k_F^4}{\hbar^2}\frac{\sin(2 k_F R) - 2 k_F R \cos(2 k_F R)}{(2 k_F R)^4},
\end{equation}
where $k_F$ is the Fermi wavevector, $R$ is the separation between bosons, and $g \propto 1/k_F$ is the interspecies interaction strength at unitarity.

In the following we describe how we obtain the interspecies interaction strength $g$ in the presence of a Fermi sea, and derive $g \propto 1/k_F$ at unitarity. The interaction strength $g$ is given by the leading order of the contact interaction strength $g_0 = (\frac{m_r}{2\pi\hbar^2 a}-\frac{2m_r}{(2\pi)^3}\int \diff^3\bm{k}\frac{1}{k^2})^{-1}$ via a regularization procedure \cite{MacNeill2011}. In a degenerate Fermi gas, the momentum integral starts from $k_F$ due to Pauli blocking, which gives $g = (\frac{m_r}{\hbar^2}(\frac{1}{2\pi a}+\frac{k_F}{\pi^2}))^{-1}$. At unitarity $1/a = 0$, we obtain $g = \frac{\pi^2\hbar^2}{m_r k_F}$, which gives $V_b \approx 3.2\frac{\hbar^2 k_F^2}{2m_f}$ for the mass ratio in our system. 

\subsection{Systematic errors}
In Fig.~2, the error bars in the horizontal direction are a combination of the error resulting from the magnetic field instability of \SI{30}{mG} and the uncertainty of the model to determine the interspecies scattering length from the magnetic field as discussed in previous sections. The vertical error bars contain the systematic error from the temperature discrepancy between Na and K and the uncertainty of the trapping frequencies.

In Fig.~3, $T$ is the average temperature for both species. The vertical error bars include the error of the fit and a systematic error from the temperature discrepancy between the Na and K clouds. As a result of this temperature mismatch between the K and Na cloud, a systematic error in lambda of 30\% was estimated.

\newpage

\end{document}